\documentclass[lettersize,journal]{IEEEtran}
\usepackage{amsmath,amsfonts}

\usepackage[algo2e,ruled,vlined,linesnumbered]{algorithm2e}
\SetInd{0em}{0.5em}
\SetKwComment{Comment}{/\!\!/}{}

\usepackage{array}
\usepackage{textcomp}
\usepackage{stfloats}
\usepackage{url}
\usepackage{verbatim}
\usepackage{graphicx}
\usepackage{cite}
\usepackage{subcaption}
\usepackage{hyperref}
\hypersetup{
    colorlinks = false,
    hidelinks
}

\usepackage{pgf}
\usepackage{pgfplots}
\usepackage{tikz}
\usetikzlibrary{spy}
\pgfplotsset{compat=1.18}

\usepackage{svg}
\usepackage{epstopdf}
\usepackage{physics}
\usepackage{makecell}

\usepackage{array}
\newcolumntype{H}{>{\setbox0=\hbox\bgroup}c<{\egroup}@{}}

\usepackage[normalem]{ulem}

\usepackage{format} 

\definecolor{burgundy}{rgb}{0.5, 0.0, 0.13}
\newcommand{\revise}[1]{\textcolor{black}{#1}}

\begin{document}
\title{GroupCDL: Interpretable Denoising and Compressed Sensing MRI via Learned Group-Sparsity and Circulant Attention}

\author{
Nikola Janju\v{s}evi\'{c}$^{1,3}$, 
Amirhossein Khalilian-Gourtani$^{2}$, 
Adeen Flinker$^{2}$, 
Li Feng$^{3}$, 
Yao Wang$^{1}$
\thanks{
$^1$New York University Tandon School of Engineering, 
Electrical and Computer Engineering Department, 
Brooklyn, NY 11201, USA.
}
\thanks{
New York University Grossman School of Medicine, 
\{%
$^2$Neurology Department, 
$^3$Radiology Department%
\},
New York, NY 10016, USA.
}
\thanks{Please send all correspondence regarding to this manuscript to N. Janju\v{s}evi\'{c} (email:nikola@nyu.edu).}}

\markboth{\LaTeX ~ Template 2024}%
{Shell \MakeLowercase{\textit{et al.}}: A Sample Article Using IEEEtran.cls for IEEE Journals}

\maketitle

\begin{abstract}
    Nonlocal self-similarity within images has become an
    increasingly popular prior in deep-learning models. 
    Despite their successful image restoration performance, such models remain largely
    uninterpretable due to their black-box construction. 
    Our previous studies have shown that interpretable
    construction of a fully convolutional denoiser (CDLNet), with performance on par with \soa
    black-box counterparts, is achievable by unrolling a convolutional dictionary learning algorithm.
    In this manuscript, we seek an interpretable construction of a convolutional
    network with a nonlocal self-similarity prior that performs on par with
    black-box nonlocal models. We show that such an architecture can be
    effectively achieved by upgrading the $\ell_1$ sparsity prior (soft-thresholding) of 
    CDLNet to an image-adaptive group-sparsity prior (group-thresholding). 
    The proposed learned group-thresholding makes use of nonlocal attention to perform
    spatially varying soft-thresholding on the latent representation.
    To enable effective training and inference on large images with global artifacts, 
    we propose a novel \emph{circulant-sparse attention}.
    We achieve competitive natural-image denoising performance
    compared to black-box nonlocal DNNs and transformers. 
    The interpretable construction of our network allows for a straightforward
    extension to Compressed Sensing MRI (CS-MRI), yielding \soa performance.
    Lastly, we show robustness to noise-level mismatches between training and inference 
    for denoising and CS-MRI reconstruction.
\end{abstract}


\section{Background and Introduction}
\IEEEPARstart{N}{onlocal} self-similarity (NLSS) of natural images has proven to
be a powerful signal prior for classical and deep-learning based image restoration. 
However, \soa NLSS deep-learning methods are widely constructed as black-boxes, 
often rendering their analysis and improvement beholden to trial and error.
Additionally, current implementations of the NLSS prior in deep-learning separately process 
overlapping image windows, falsely neglecting the dependency between these
overlaps. Here, we address these two shortcomings of nonlocal deep
neural networks (DNNs) from the perspective of interpretable architecture
design and sparse array arithmetic. This construction allows us to extend the
formulation to compressed sensing magnetic resonance imaging (CS-MRI) without
use of tricks common in adapting black-box methods to inverse-problems.

A growing literature of DNNs, derived as direct parameterizations of classical
image restoration algorithms, perform on par with \soa black-box fully
convolutional neural networks, without employing common deep-learning tricks
(such as batch-normalization, residual learning, and feature domain
processing). \revise{This interpretable construction has been shown to be instrumental
in obtaining parameter and dataset efficiency\cite{janjusevicCDLNet2022,
lecouat2020nonlocal, janjusevicGDLNet2022, Simon2019, Scetbon2019DeepKD, Gu2022}
}. Our
previous work, CDLNet \cite{janjusevicCDLNet2022}, introduced a unique
interpretable construction, based on convolutional dictionary learning, and
achieved novel robustness to mismatches in observed noise-level during training
and inference. In this work, we incorporate the NLSS prior into CDLNet and
demonstrate the first instance of an interpretable network bridging the
performance gap to \soa nonlocal black-box methods for image denoising.
Additionally, the extension of our method to CS-MRI outperforms existing \soa
methods.

Nonlocal layers in DNNs attempt to model long-range image dependencies by
computing pixel-wise self-similarity metrics. To tame the quadratic
computational complexity of this operation, image-restoration DNNs generally
rely on computing dense similarities on small overlapping patches of the input,
which are processed independently by the layer/network and subsequently
averaged on overlapping regions to produce the final output
\cite{lecouat2020nonlocal, liu2018non, Liang_2021_ICCV}. \revise{Naturally, such
patch-based dense attention (PbDA) may incur a runtime penalty by redundant overlap
processing and a restoration penalty due to the disregard for the correlation
among these overlapping regions \cite{Simon2019}}. 

\revise{We ultimately achieve the incorporation of the NLSS prior into CDLNet by two
orthogonal but complementary contributions: learned group-thresholding (GT) and
circulant-sparse attention (CircAtt), which together we call GroupCDL. The
former encodes the functional form of NLSS via a spatially-varying,
image-adaptive, soft-thresholding layer, which is dictated by an adjacency
matrix $\ADJMAT$. The latter concerns the formation of $\ADJMAT$ in a
computationally efficient and differentiable manner, all while maintaining the
translation-equivariance inductive bias of convolutional sparse-coding. Not
only does CircAtt address the modeling issues of PbDA, it also allows for
seamless training and inference on large images, enabling more effective
restoration of global degradation such as aliasing-artifacts encountered in
CS-MRI.}

Previous work used a patch-based group-sparsity prior in an
interpretably constructed DNN \cite{lecouat2020nonlocal}. However, this approach
did not achieve competitive performance with black-box nonlocal DNNs. 
In contrast, we enforce pixel-wise group-sparsity of a latent convolutional
sparse representation, with nonlocal attention performed in a
compressed subband domain to aid performance and inference speed.
We also propose the novel CircAtt operation, together achieving denoising
performance on par with the \soa methods.  

We highlight the following contributions:
\begin{itemize}
    \item \revise{a novel and efficient nonlocal self-similarity operation (CircAtt)
        which addresses the modeling and computational shortcomings of common
        patch-based dense attention (PbDA).}
    \item \revise{a novel learned group-thresholding operation, which encodes the functional form of the group-sparsity prior in a differentiable manner, and
        utilizes a reduced channel dimension of the latent space to
        achieve competitive inference speeds.}
    \item \revise{an interpretable nonlocal CNN with competitive 
        denoising and CS-MRI performance to \soa black-box models, and novel robustness to noise-level mismatches between training and inference.}
    \item an open-source implementation\footnote{ 
        \href{https://github.com/nikopj/GroupCDL}{https://github.com/nikopj/GroupCDL}, \\
        \href{https://github.com/nikopj/GroupCDL}{https://github.com/nikopj/CirculantAttention.jl}.
    } in the Julia programming language \cite{julia}.
\end{itemize}

The remainder of the manuscript is organized as follows: in Section
\ref{sec:prelim}, we introduce the mathematics and notation behind
classical convolutional dictionary learning and group-sparse representation. We
also provide context for related black-box and interpretable deep-learning
methods. In Section \ref{sec:method}, we introduce our nonlocal
CNN derived from group-sparse convolutional dictionary learning, dubbed GroupCDL. We additionally
introduce GroupCDL's extension for CS-MRI reconstruction and detail our novel CircAtt operation.
In Section \ref{sec:results}, we show experimental results that compare GroupCDL
to \soa deep learning methods for both denoising and CS-MRI reconstruction.

\section{Preliminaries and Related Work} \label{sec:prelim}
\begin{table}[tb]
\caption{Notation}
\centering
\resizebox{\linewidth}{!}{%
\begin{tabular}{|l|l|} \hline
\multirow{2}{*}{$\x \in \R^{NC}$} & a vector valued image with $N=N_1\times N_2$ pixels, 
\\ & and vectorized channels, $\x =[ \x_1^T,\, \cdots,\, \x_C^T ]^T$. \\
\hline
$\x_c \in \R^N$ & the $c$-th subband/feature-map/channel of $\x$. \\
\hline
$\x[n] \in \R^C$ & the $n$-th pixel of $\x$, $n \in [1, \, N]$. \\
\hline
$\x_c[n] \in \R$ & the $n$-th pixel of the $c$-th channel of $\x$. \\
\hline
$\vec{n}\in [1, N_1] \times [1,N_2]$ & the spatial coordinates of the $n$-th pixel of $\x$. \\
\hline
$\bu \circ \bv \in \R^N$ & the element-wise product of two
vectors. \\
\hline
\multirow{2}{*}{$\bD \in \R^{NC \times QM}$} & a 2D $M$ to $C$
channel synthesis convolution \\ & operator with stride $s_c$, where $Q=N/s_c^2$.\\
\hline
\multirow{2}{*}{$\bD^T \in \R^{QM \times NC}$} & a 2D $C$ to $M$
channel analysis convolution \\ & operator with stride $s_c$, where $Q=N/s_c^2$.\\
\hline
$\bU \in \R^{Q\times N}$ & a $Q \times N$ matrix with elements $\bU_{ij} \in \R$. \\
\hline
$\bU_{i:} \in \R^N, ~ \bU_{:j} \in \R^Q$ & the $i$-th row, $j$-th column of matrix $\bU$. \\
\hline
\multirow{2}{*}{$\bU \otimes \bV \in \R^{QM \times NC}$} & Kronecker product of $\bU \in
\R^{Q\times N}$ and $\bV \in \R^{M\times C}$, \\
& i.e. the block matrix with $\bV$s scaled by $\bU_{ij} ~ \forall ~ i,j$.\\
\hline
$\IDMAT_N \in \R^{N \times N}$ & the $N$ by $N$ identity matrix. \\
\hline
\multirow{2}{*}{$\y = \cwise{\bU}\x \equiv (\IDMAT_C \otimes \bU)\x$} & the matrix $\bU$ applied channel-wise, \\
& i.e. $\y_c = \bU\x_c \in \R^Q, ~ \forall ~ 1 \leq c \leq C$. \\
\hline
\multirow{2}{*}{$\y = \pwise{\bV}\x \equiv (\bV \otimes \IDMAT_N)\x$} & the matrix $\bV$ applied pixel-wise,\\
& i.e. $\y[i] = \bV\x[i] \in \R^M, ~ \forall ~ 1 \leq i \leq N$. \\
\hline
\multirow{2}{*}{$\bS \in \BCCB_W^{N \times N}$} & a real-valued $N\times N$ matrix with \\ 
& BCCB sparsity pattern of windowsize $W$.  \\
\hline
\end{tabular}}
\label{tab:notation}
\end{table}
\subsection{Dictionary Learning and Group-Sparse Representation}
We consider the observation model of additive white Gaussian noise (AWGN),
\begin{equation} \label{eq:awgn}
    \y = \x + \boldsymbol{\nu}, \quad \textnormal{where}\quad \boldsymbol{\nu}\sim\N(\boldsymbol{0}, \sigma^2 \IDMAT).
\end{equation}
Here, the ground-truth image $\x \in \R^{NC}$ is contaminated with AWGN of
noise-level $\sigma$, resulting in observed image $\y \in \R^{NC}$. For
convenience and clarity of notation, we denote images in the vectorized form,
and any linear operation on an image as a matrix vector multiplication (see
Table \ref{tab:notation} for details). In implementation, fast algorithms are
used and these matrices are not actually formed, except when explicitly
mentioned. 

We frame our signal-recovery problem in terms of a (given)
$s_c$-strided convolutional dictionary $\bD \in \R^{NC \times QM}$, with $Q=N/s_c^2$, i.e. the
columns of $\bD$ are
formed by integer translates of a set of $M$ (vectorized) 2D convolutional
filters, each having $C$ channels.
We assume $\exists \,
\z \in \R^{QM}\, \st \, \x \approx \bD\z$. 
The rich works of sparse-representation and compressed-sensing
provide guarantees based on assumptions of sparsity in $\z$ and regularity on
the columns of $\bD$ \cite{Mallat}. We refer to $\z$ as our sparse-code,
latent-representation, or subband-representation of $\x$.

A popular classical paradigm for estimating $\x$ from an observed
noisy $\y$ is the Basis Pursuit DeNoising (BPDN) model, 
\begin{equation} \label{eq:bpdn}
\underset{\z}{\mathrm{minimize}} ~ \frac{1}{2}\norm{\y -\bD\z}_2^2 + \lambda \psi(\z),
\end{equation}
where $\psi : \R^{QM} \rightarrow \R_+$ is a chosen regularization function. The
Lagrange-multiplier term $\lambda > 0$ provides a trade-off between satisfying
observation consistency and obeying the prior-knowledge encoded by $\psi$. A
popular approach to solving \eqref{eq:bpdn} is the proximal-gradient method
(PGM) \cite{Beck2009}, involving the {\it proximal-operator} of $\psi$, defined as  
\revise{
\begin{equation} \label{eq:prox}
    \prox_{\tau \psi}(\bv) \coloneqq \argmin_{\z} \tau \psi(\z) +
    \frac{1}{2}\norm{\z - \bv}_2^2, \quad \tau > 0.
\end{equation}
}
PGM can be understood as a fixed point iteration involving the iterative application
of a gradient-descent step on the observation consistency term of \eqref{eq:bpdn} followed by 
application of the proximal operator of $\psi$,
\begin{equation} \label{eq:pgm}
\z^{(k+1)} = \prox_{\tau \psi}(\z^{(k)} - \eta \bD^T(\bD\z^{(k)} - \y)),
\end{equation}
where $\tau =\eta\lambda$,  and $\eta > 0$ is a step-size parameter.

When $\psi$ is the sparsity-promoting $\ell_1$-norm, the proximal operator is
given in closed-form by element-wise soft-thresholding \cite{Beck2009},
\begin{equation} \label{eq:ST}
\ST_\tau(\z) = \z \circ \left(1 - \frac{\tau}{\abs{\z}}\right)_+,
\end{equation}
where $(\cdot)_+$ denotes projection onto the positive orthant $\R_+$. 
The resulting PGM iterations are commonly referred
to as the Iterative Soft-Thresholding Algorithm (ISTA) \cite{Beck2009}.

More sophisticated priors ($\psi$) can be used to
obtain better estimates of our desired ground-truth image by exploiting
correlations between ``related" image-pixels. One such prior is
{\it group-sparsity}, 
\begin{align} \label{eq:group_sparse}
    \psi(\z) &= \sum_{\substack{m=1 \\i=1}}^{M, Q}
    \sqrt{\sum_{j=1}^Q \ADJMAT_{ij} \z_m[j]^2} 
             = \norm{ \sqrt{(\IDMAT_M \otimes \ADJMAT) \z^2} }_1, 
\end{align}
where $\ADJMAT \in \R_+^{Q\times Q}$ is a row-normalized adjacency matrix (i.e.
$\norm{\ADJMAT_{i:}}_1 = 1$), and $\cdot^2$ and $\sqrt{\cdot}$ are taken element-wise.
The $i$th adjacency row ($\ADJMAT_{i:}$) measures closeness between the $i$th
latent pixel ($\z[i]$) and every other latent pixel.
Hence, group-sparse regularization may be understood as encouraging similar
latent-pixels to share the same channel-wise sparsity pattern, and has been
shown to improve denoising performance under classical patch-based sparse
coding methods \cite{mairal2009non}, as well as recent interpretably
constructed DNNs \cite{lecouat2020nonlocal}. \revise{Group-sparsity is closely related
to the joint-sparsity prior, and may be considered as joint-sparsity of codes
throughout an image belonging to the same group/cluster as defined by
$\ADJMAT$, instead of joint-sparsity of color-coefficients
\cite{janjusevicCDLNet2022} or MRI coil-coefficients \cite{Murphy2012TMI} of the
same pixel location.}

\revise{When $\ADJMAT$ is a (row-normalized) binary symmetric adjacency matrix, the
sparse-codes are separated into clusters and the group-sparsity proximal
operator is given by block-thresholding of each group \cite{Usman2011MRM,
janjusevicCDLNet2022}, or in this context known as group-thresholding (GT),
\begin{equation} \label{eq:GT}
    \begin{aligned} 
        \GT_{\tau}(\z; \ADJMAT) &= \z \circ \left( 1 - \frac{\tau}
            {\sqrt{(\IDMAT_M \otimes \ADJMAT)\z^2}} \right)_+.
    \end{aligned} 
\end{equation}}

\revise{Motivated by \cite{lecouat2020nonlocal}, we relax the binary symmetric
requirement to obtain a generalized group-thresholding, using the functional
form of \eqref{eq:GT} with an arbitrary $\ADJMAT \in \R_+^{Q\times Q}$. This
relaxation makes GT only an approximate solution to
\eqref{eq:prox}, but maintains the desirable property of GT \eqref{eq:GT}
reducing to element-wise soft-thresholding \eqref{eq:ST} when $\ADJMAT$ is the
identity matrix. Note that this classical group-thresholding \eqref{eq:GT}
assumes $\ADJMAT$ is given, and in practice it is often pre-computed from the
input image.}

\revise{From the perspective of deep-learning,
GT involves an ``attention" operation: a signal transformation via an adjacency
matrix $\ADJMAT$. Black-box attention networks (e.g. transformers)
use attention to generate features for the next layer via matrix-vector
multiplication given an adjacency matrix, ex. ${(\IDMAT_M \otimes \ADJMAT)\z}$. 
In contrast, GT performs attention to obtain an adjacency weighted signal
energy which informs a spatially varying soft-thresholding, i.e.
group-thresholding may be rewritten as,}
\revise{
$$
\GT_\tau(\z; \ADJMAT) = \ST_{\frac{\tau}{\bxi}\abs{\z}}(\z), \quad \bxi = \sqrt{(\IDMAT_M \otimes \ADJMAT)\z^2}. 
$$}

The BPDN \eqref{eq:bpdn} problem can be made more expressive by opting to learn
an optimal dictionary from a dataset of noisy images $\D = \{\y\}$. \revise{We express
the (convolutional) dictionary learning problem as,
\begin{equation} \label{eq:dict_learn}
\underset{\{\z\}, \bD \in \mathcal{C}}{\mathrm{minimize}} ~ 
\sum_{\y \in \mathcal{D}} \frac{1}{2}\norm{\y -\bD\z}_2^2 + \lambda
\psi(\z),
\end{equation}
where constraint set $\mathcal{C} = \{ \bD \, : \, \norm{\bD_{:j}}_2^2 \leq 1 ~
\forall \, j \}$ ensures that the regularization term is not rendered useless by
an arbitrary scaling of latent coefficients \cite{janjusevicCDLNet2022, mairal2009online}.} Solving \eqref{eq:dict_learn}
generally involves alternating sparse-pursuit (ex. \eqref{eq:pgm}) and a
dictionary update with fixed sparse-codes (ex. projected gradient descent)
\cite{mairal2009online}. 

\subsection{Unrolled and Dictionary Learning Networks}
%
Approaches in \cite{ongie2020deep, Gilton2019, deqWillet2021} explore the
construction of DNNs as unrolled proximal gradient descent machines with
proximal-operators that are implemented by a black-box CNN, learned end-to-end. 
Although these methods contribute to more principled DNN architecture design in
image-processing, their use of black-box neural networks, such as UNets
\cite{unet} and ResNets \cite{he2016deep}, ultimately side-step the goal of
full interpretability. In contrast, our previous work CDLNet
\cite{janjusevicCDLNet2022} introduces a CNN as a direct parameterization of
convolutional PGM \eqref{eq:pgm} with an $\ell_1$ sparsity prior.
In this manuscript, we extend the formulation and direct parameterization of
CDLNet by introducing a novel implementation of the group-sparsity prior,
embodied in the proposed GroupCDL architecture (see Section \ref{sec:method}).
We also show that the noise-adaptive thresholding of CDLNet, derived from BPDN
\eqref{eq:bpdn}, extends to GroupCDL in both image denoising and joint
denoising and CS-MRI (see Section \ref{sec:exp:csmri}).

Zheng et. al \cite{Zheng_2021_CVPR} propose a DNN architecture based on a
classical dictionary learning formulation of image denoising. However, this
network heavily employs black-box models such as UNets \cite{unet} and
multi-layer perceptrons (MLPs). Our proposed method differentiates itself by
using a direct parameterization of variables present in the classical proximal
gradient method \eqref{eq:pgm} with a group-sparsity regularizer
\eqref{eq:group_sparse}. This construction offers an alternative to the use of
black-boxes, yielding great learned parameter count efficiency, novel
generalization capabilities, and easy extensibility. 

Directly parameterized dictionary learning networks \cite{janjusevicCDLNet2022,
janjusevicGDLNet2022, lecouat2020nonlocal, Sreter2018, Simon2019,
Scetbon2019DeepKD} have gained some popularity in recent years due to their
simple design and strong structural similarities to popular ReLU-activation
DNNs. This connection was first established by the seminal work of Gregor et.
al \cite{Gregor2010} in the form of the Learned Iterative Shrinkage
Thresholding Algorithm (LISTA). Here, we build upon this literature by
proposing a learned approximate proximal operator for the convolutional
dictionary learning formulation of a DNN, derived from a group-sparsity prior
(see Section \ref{sec:groupcdl}). We demonstrate that such a network can
compete well with and outperform \soa methods, without sacrificing
interpretability. 

Lecouat et. al \cite{lecouat2020nonlocal} propose a nonlocal CNN derived from a
patch-based dictionary learning algorithm with a group-sparsity prior, dubbed
GroupSC. It is well established that the independent processing of image
patches and subsequent overlap and averaging (patch-processing) is inherently
suboptimal to the convolutional model, due to the lack of consensus between
pixels in overlapping patch regions \cite{Simon2019}. Our method is in-part
inspired by GroupSC, but is adapted to the convolutional sparse coding with
groups defined over sliding windows via CircAtt. 

\subsection{Nonlocal Networks}
\revise{The nonlocal self-similarity prior in image-restoration DNNs is commonly
formulated with patch-based dense attention (PbDA) to manage quadratic complexity \cite{liu2018non,
lecouat2020nonlocal}.} Patch overlap is often used to ensure that
artifacts do not occur on local-window boundaries. Despite many such networks 
being formulated as CNNs, their patch-based inference ultimately diminishes the
powerful shift-invariance prior and increases computational cost due to
additional processing of overlapping regions (see Section
\ref{sec:slidingwindow}). 

\revise{Liu et. al \cite{liu2021Swin} proposed an alternate form of PbDA via the
Shifted-Window Vision Transformer architecture (SwinViT), demonstrated on image
classification, segmentation, and detection tasks. Their so-called Swin
attention mechanism makes use of patching without overlap, using alternating
``shifted" patch boundaries between PbDA layers.} This allows some cross over of
information of neighboring patches \cite{liu2021Swin}. Liu et. al's ablation
studies considered use of a sliding-window attention mechanism, similar to our
CircAtt, but ultimately abandoned it due to similar performance and slower
inference compared to Swin attention on their examined tasks. 
Liang et. al \cite{Liang_2021_ICCV} (SwinIR) adapted the SwinViT
architecture for image denoising and other restoration tasks with local
degradation operators (ex. super-resolution, JPEG artifact removal).
Our proposed method re-examines sliding-window attention in the context of
image restoration and shows favorable performance over SwinIR in terms of
learned parameter count efficiency.

Zamir et. al \cite{Zamir2021Restormer} proposed Restormer, a multi-resolution
transformer model for image restoration. They demonstrated results on image
denoising and local degradation operator tasks. Restormer's core (transformer)
block does not use nonlocal attention, but instead computes similarities
between entire feature maps (known as ``transposed attention"). In contrast,
the proposed GroupCDL method operates on a single resolution and implements
nonlocal attention.

To correctly account for dependencies between neighboring local windows, we
propose a sliding-window NLSS (CircAtt), enabled by sparse matrix arithmetic.
To the best of our knowledge, we are the first to examine a sliding-window
attention mechanism for image restoration. Recent works have proposed other
so-called ``sparse attention" mechanisms, however, they have either not been in
the context of image restoration \cite{child2019sparsetransformer}, not
employed a sliding-window \cite{dao2022flashattention}, or have employed a
complicated hashing algorithm to exploit extremely long-range dependencies
\cite{Mei_2021_CVPR}.

\section{Proposed Method} \label{sec:method}
\begin{figure*}[thb]
    \centering
    \includegraphics[width=\textwidth]{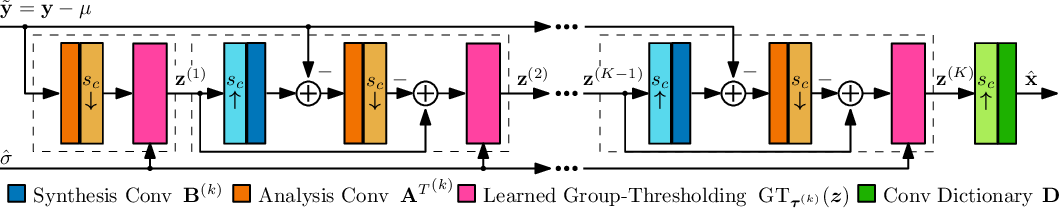}
    \caption{\revise{
        The GroupCDL Architecture for Denoising: the Soft-Thresholding of
        CDLNet is replaced with a Learning Group-Thresholding ($\GT$) operator,
        encoding a nonlocal self-similarity prior via Circulant-Sparse
        Attention (CircAtt). An expanded view of Learned $\GT$ is given in Figure \ref{fig:GT_block}.}
}
    \label{fig:arch}
\end{figure*}
\begin{figure}[thb]
    \centering
    \vspace*{-1em}
    \includegraphics[width=0.85\columnwidth]{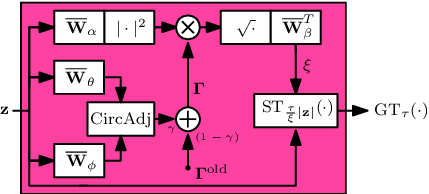}
    \caption{\revise{
        Expanded block diagram of Learned Group-Thresholding
        \eqref{eq:GTlearned} with Circulant-Sparse Attention from the GroupCDL
        Architecture (Fig. \ref{fig:arch}). Learned transforms
        ($\bW_{\theta,\phi,\alpha,\beta}$) are used to reduce computational
        complexity by computing and applying sparse attention a reduced number
        of channels. $\mathrm{CircAdj}$ refers to row-softmax
        normalization of a Circulant-Sparse similarity matrix
        \eqref{eq:GTlearned}. The last computed adjacency matrix
        $(\ADJMAT^{\mathrm{old}})$ is kept to combine with its new computation
        via learned parameter $\gamma \in [0,1]$.}
    }
    \vspace*{-1.5em}
    \label{fig:GT_block}
\end{figure}
\revise{Our proposed framework, GroupCDL, derives nonlocal attention from the
group-sparsity prior \eqref{eq:group_sparse}. 
GroupCDL adapts the classical group-thresholding operator \eqref{eq:GT} to our
proposed learned group-thresholding \eqref{eq:GTlearned}, all in a
computationally efficient manner via circulant-sparse self-similarity and
attention computations. We first introduce the proposed GroupCDL architecture
for the problem of image restoration under the AWGN model \eqref{eq:awgn}
(Section \ref{sec:groupcdl}). AWGN is a popular and successful model for camera
noise following white-balance and gamma-correction \cite{Khashabi2014}.
Denoising also serves as a fundamental building block (ex. proximal operators)
and subproblem in various inverse-problem approaches \cite{Gilton2019}. Hence,
our proposed method is readily adapted to applications beyond camera image
denoising.
We demonstrate this fact in Section \ref{sec:csmri} by extending the proposed
method to Compressed Sensing MRI.}
In Section \ref{sec:circatt}, we detail our Circulant-Sparse Adjacency and Attention mechanisms (Section
\ref{sec:circatt}), which are integral to implementing learned group-thresholding on
convolutional sparse codes, in addition to enabling unified training and
inference on large images. 
Lastly, Sections \ref{sec:slidingwindow} and \ref{sec:dpa} give 
discussion on the relation between our interpretably constructed nonlocal
network and common black-box approaches.

\subsection{The GroupCDL Architecture} \label{sec:groupcdl}
\begin{figure}[thb]
    \centering
    \begin{subfigure}{\linewidth}
        \centering
        \includegraphics[width=\linewidth]{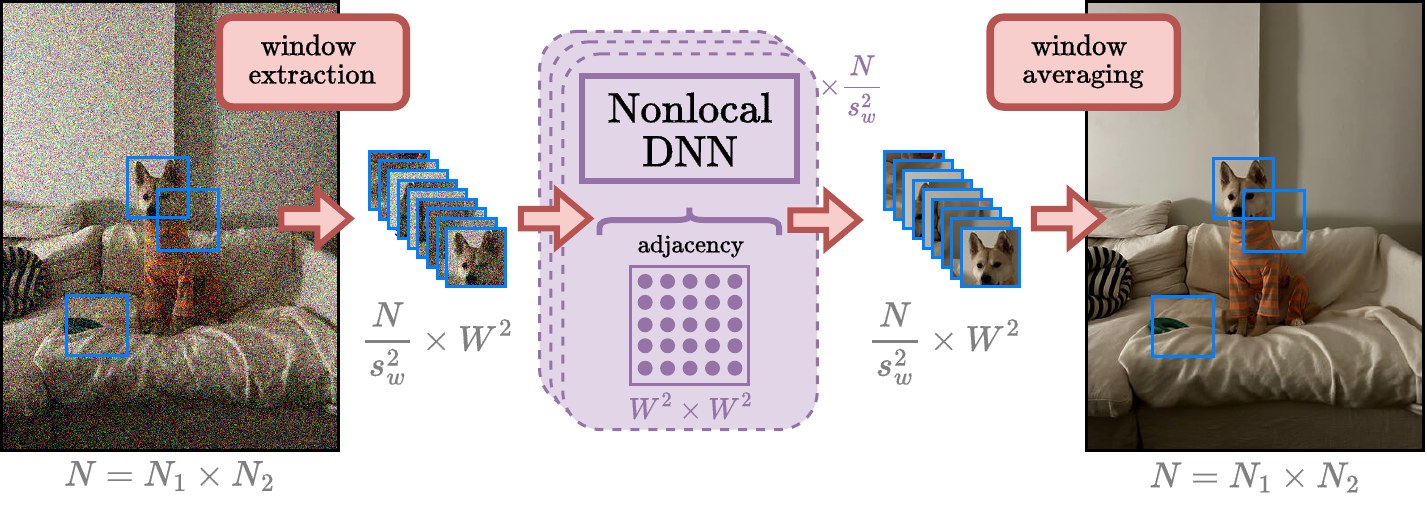}
        \caption{Patch-based Dense Attention (PbDA)}
        \vspace*{1em}
        \label{fig:HT:dense}
    \end{subfigure}
    \begin{subfigure}{\linewidth}
        \centering 
        \includegraphics[width=0.9\linewidth]{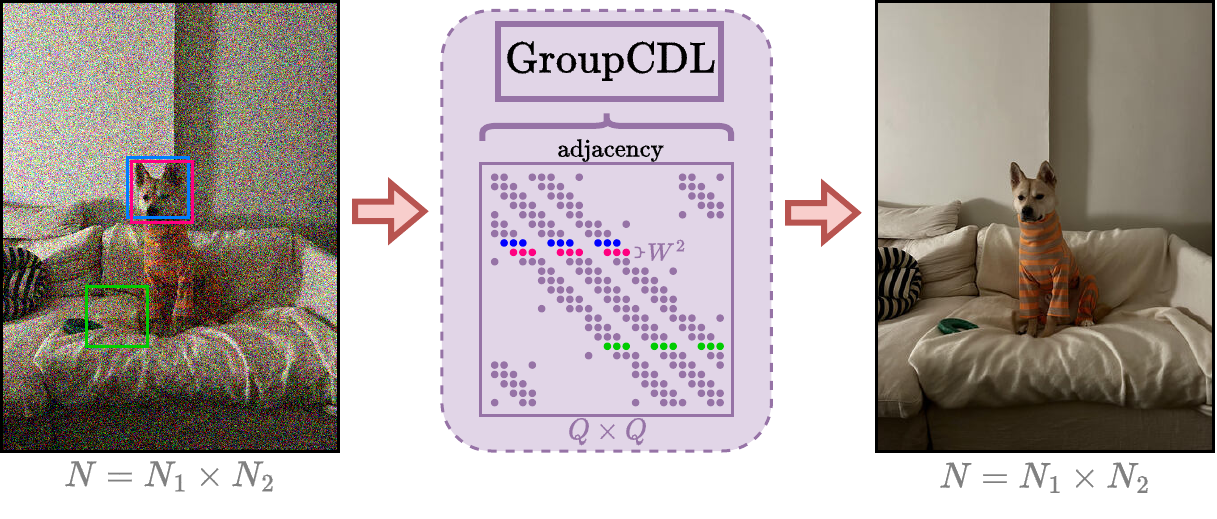}
        \caption{Circulant-Sparse Attention (CircAtt)}
        \label{fig:HT:circ}
    \end{subfigure}
    \caption{
        (a) To enable dense attention, the input image is divided into
        overlapping windows (of size $\WINSZ\times \WINSZ$ and with
        window-stride $s_w \times s_w$), processed independently via a DNN with
        dense self-attention. The denoised windows are then placed in their
        original positions and averaged on their overlaps. (b) CircAtt
        processes the entire image jointly in a single forward pass. The
        adjacency matrix has a block-circulant with circulant blocks (BCCB)
        sparsity pattern, where the number of non-zeros in each row/column is
        at most $\WINSZ^2$. The adjacency matrix is computed in the subband
        domain, with spatial dimension $Q=N/s_c^2$, where $s_c$ is the
        convolution stride. Hence, the effective image-domain window-size is
        $s_c\WINSZ \times s_c\WINSZ$.
    }
    \label{fig:HT}
\end{figure}

\revise{
We propose a neural network architecture as a direct parameterization of PGM
\eqref{eq:pgm} on the convolutional BPDN problem with a group-sparsity prior,
dubbed GroupCDL. The GroupCDL architecture is equivalent to replacing the
CDLNet \cite{janjusevicCDLNet2022} architecture's soft-thresholding
\eqref{eq:ST} with a learned group-thresholding \eqref{eq:GTlearned}, as shown in Figures \ref{fig:arch},
\ref{fig:GT_block}. Given a noisy input image $\y \in \R^{NC}$, the GroupCDL network is defined as,
\begin{equation} \label{eq:GroupCDL}
\begin{gathered}
\z^{(0)} = \bm{0}, \quad \text{for } ~ k=0,1,\dots, K-1,\\
\z^{(k+1)} = \GT_{\boldsymbol{\tau}^{(k)}}\left(\z^{(k)} - {\bm{A}^{(k)}}^T(\bm{B}^{(k)}\z^{(k)} - \tilde{\y}) \right), \\
\btau^{(k)} = \btau^{(k)}_0 + \hat{\sigma}\btau^{(k)}_1, \quad \hat{\x} = \bm{D}\z^{(K)} + \mu.
\end{gathered}
\end{equation}
Where $\mu = \mathrm{mean}(\y)$ and $\tilde{\y} = \y - \mu$.
Here, ${\bA^T}^{(k)}, \bB^{(k)}$ are 2D ($C$ to $M$ channel,
stride-$s_c$) analysis and ($M$ to $C$ channel, stride-$s_c$) synthesis
convolutions, respectively. $\bD$ is our 2D ($M$ to $C$ channel, stride-$s_c$)
synthesis convolutional dictionary. For an input noisy image $\y \in \R^{NC}$,
our latent representation is of the form $\z \in \R^{QM}$, where {$Q=N/s_c^2$}.
}
\revise{
In classical group-sparsity, a binary symmetric similarity matrix is generally pre-computed before optimization.
To enable differentiability and increase expressivity of GroupCDL, we propose to compute and update 
a soft similarity matrix throughout the network inside of a learned group-thresholding operator as follows,
\begin{equation} \label{eq:GTlearned}
    \begin{gathered}
        \GT_{\btau}(\z) = \z \circ \left(1 - \frac{\btau} { \pwise{\bW^T}_{\beta} \sqrt{(I_{M_h} \otimes \ADJMAT)(\pwise{\bW}_{\alpha} \z)^2}} \right)_+ \\
        \ADJMAT = \rowsoftmax(\SIMMAT) \\
        \SIMMAT_{ij} = \begin{cases}
            -\frac{1}{2} \norm{\tfrac{1}{\brho} \circ (\bW_\theta \z[i] - \bW_\phi \z[j])}_2^2, & \norm{\vec{i} - \vec{j}}_{\infty} \leq \tfrac{1}{2}W \\
            -\infty, & \text{else}
        \end{cases},
    \end{gathered}
\end{equation}
where $\rowsoftmax$ refers to a row-wise softmax operation.
Learned group-thresholding \eqref{eq:GTlearned} accepts a convolutional sparse code $\z$,
and first computes a similarity matrix $\SIMMAT \in \BCCB_W^{Q\times Q}$ with pixel-wise transforms 
$\bW_\theta, \bW_\phi \in \R^{M_h \times M}$. To reduce computational
complexity, these similarity computations are only performed between pixels $i$
and $j$ of the sparse code that are within a $W\times W$ window of each other, with circular boundary conditions.
The resulting sparsity pattern of $\SIMMAT$ is block-circulant with circulant blocks (BCCB), i.e. that of a 2D sliding window (see Fig. \ref{fig:HT:circ}).
We dub this Circulant-Sparse Distance-Similarity (Alg. \ref{alg:circdist}) (further details in Section \ref{sec:circatt}). 
}

\revise{
The circulant-sparse adjacency matrix $\ADJMAT$ is obtained by a row-softmax normalization $\SIMMAT$.
$\ADJMAT$ is then applied channel-wise to obtain a spatially varying threshold
value. However, different to \eqref{eq:GT}, we employ learned pixel-wise
transforms $\bW_\alpha \in \R^{M_h\times M}, \bW_\beta \in \R_+^{M_h\times M}$
before and after the attention computation. In practice, we choose to reduce
the channel-dimension for attention, i.e. $M_h << M$, which has the effect of
greatly reducing the computational cost of attention. In Section
\ref{sec:exp:ablation}, we show that these savings do not harm denoising performance.
}

\revise{
To further reduce computational complexity, we only compute similarity every $\Delta K$
layers. To ensure smooth adjacency matrix updates, a convex combination of the
computed adjacency and the adjacency of the previous layer is employed via a
learned parameter $\gamma \in [0,1]$,
\begin{equation} \label{eq:adj_update}
    \ADJMAT^{(k+1)} = \gamma \ADJMAT^{(k)} + (1-\gamma) \ADJMAT^{(k-1)}.
\end{equation}
}

\revise{
Figure \ref{fig:GT_block} represents the proposed learned group-thresholding in
block-diagram form. To prevent over-fitting, we share the pixel-wise transforms
of learned $\GT$ and the combination parameter $\gamma$ in all layers of the
network. However, we untie the similarity parameter $\brho^{(k)} \in \R^{M_h}$
so that the network may learn to become more confident in its similarity
computations as more noise has been removed deeper into the network.
Altogether, the parameters of GroupCDL are ${\Theta =
\{\bD, \, \gamma, \, \bW_{\theta, \phi, \alpha, \beta}, \,
\{{\bA^T}^{(k)}, \,\bB^{(k)}, \, \btau_0^{(k)}, \btau^{(k)}_1, \brho^{(k)}
\}_{0=1}^{K-1}\}}$.
}

\begin{figure}[bht]
    \centering
    \includegraphics[width=0.8\columnwidth]{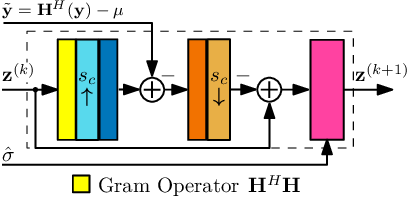}
    \caption{
        Block diagram of a single GroupCDL layer in Fig. \ref{fig:arch} extended to general linear
        inverse problem $\y = \bm{H}\x + \nu$. The layer differs from the
        denoising GroupCDL by adjoint operator preprocessing and use of the
        Gram Operator $\bm{H}^H\bm{H}$ in each layer. In the case of 
        CS-MRI, the observation operator is a masked multi-coil Fourier transform following a sensitivity map operator, 
        $\bm{H} = \cwise{\bM\bF}\bR$. Other blocks correspond with Figure \ref{fig:arch}.
    }
    \label{fig:arch_gram}
\end{figure}
\subsection{Extending the GroupCDL Architecture for CS-MRI} \label{sec:csmri}
Parallel (or multi-coil) compressed sensing magnetic resonance imaging
(CS-MRI) may in general be described by the following observation model,
\begin{equation} \label{eq:csmri-obs}
    \y = \cwise{\bM}(\cwise{\bF} \bR \x + \bnu), \quad \bnu \sim \CN(0, \pwise{\bSigma}).
\end{equation}
where $\x \in \C^N$ is the ground-truth image-domain signal and $\y \in
\C^{NC}$ is a complex-valued measurement vector with $C$-channels acquired in a
masked Fourier domain (where $C$ is the number of scanner coils). Here, $\bM = \diag(\bbm)$ is channel-wise mask operator
with binary mask $\bbm \in \{0,1\}^N$ indicating the position of acquired
Fourier domain samples. $\bF$ represents the $N$-dimensional 2D-DFT matrix,
which is applied channel-wise. The coil-sensitivities are encoded in the
sensitivity operator 
$
\bR = 
\begin{bmatrix} \diag(\br_1) & \cdots &
\diag(\br_C) 
\end{bmatrix}^T
$ 
via sensitivity map $\br \in \C^{NC}$. The
acquired samples are well modeled as being contaminated by i.i.d. 
complex additive Gaussian noise with noise-covariance matrix
$\bSigma \in \C^{C\times C}$. This observation model may be simplified by
considering a coil-whitening pre-processing transformation,
resulting in the noise following a diagonal covariance matrix of equal power in each coil, 
i.e. $\bnu \sim \CN(0, \sigma^2\IDMAT)$. After coil-whitening, the CS-MRI convolutional dictionary
BPDN functional takes the form,
\begin{equation} \label{eq:csmri-bpdn}
\underset{\z}{\mathrm{minimize}} ~ \frac{1}{2}\norm{\y -\cwise{\bM\bF}\bR\bD\z}_2^2 + \lambda \psi(\z).
\end{equation}
\revise{Note that the underbar notation indicates a matrix is applied channel-wise (see
Table \ref{tab:notation}).}
We derive the CS-MRI GroupCDL (and CS-MRI CDLNet) architectures as a directly
parameterized unrolling of PGM applied to \eqref{eq:csmri-bpdn}, employing
complex-valued filter-banks/dictionaries. \revise{Specifically,
denote the observation operator $\bH
= \cwise{\bM\bF}\bR$. We then transform the input kspace signal $\y$ to a
zero-filled reconstruction with mean-subtraction, $\tilde{\y} = \bH^H\y - \mu$,
with $\mu = \mathrm{mean}(\bH^H\y)$. We insert the observation Gram
operator $\bH^H\bH$ after synthesis convolution in each layer of GroupCDL,
\begin{equation} \label{eq:cdl_csmri}
    \z^{(k+1)} = \GT_{\boldsymbol{\tau}^{(k)}}\left(\z^{(k)} - {\bm{A}^{(k)}}^H(\bH^H\bH\bm{B}^{(k)}\z^{(k)} - \tilde{\y}) \right). 
\end{equation}
Figure \ref{fig:arch_gram} illustrates this change in a single CDLNet/GroupCDL layer.}

\begin{algorithm2e}[tbh]
\caption{\revise{Circulant-Sparse Dist-Sim {\small $\mathcal{O}(QW^2M)$}}}
\label{alg:circdist}
\DontPrintSemicolon
\begin{small}
\SetKwFunction{FFwd}{CircDistSim}
\SetKwProg{Pn}{function}{:}{\KwRet}
\SetKwFunction{FBwd}{bwd}
\Pn{\FFwd{$\bk \in \C^{QM}$, $\bq \in \C^{QM}$; $W \in \Z_+$}}{
    $\SIMMAT_{ij} = 
    \begin{cases}
        -\tfrac{1}{2}\norm{\bk[i] - \bq[j]}_2^2, &
        \norm{\vec{i} - \vec{j}}_\infty \leq \tfrac{1}{2}\WINSZ \\
        - \infty , & \text{otherwise}
    \end{cases} {\textstyle \forall ~ (i,j) \in [1,Q]^2}$\; 

    \BlankLine

    \Pn{\FBwd{$\dd\SIMMAT \in \BCCB_W^{Q\times Q}$}}{
        $\dd{\bk} = \cwise{\dd\bS}\bq - (\cwise{\dd\bS} \bOne) \circ \bk$\;
        $\dd{\bq} = \cwise{\dd\bS}^T\bk - (\cwise{\dd\bS}^T \bOne) \circ \bq$\;
        \KwRet $\dd{\bk}$, $\dd{\bq}$\;
    }
    \KwRet $\SIMMAT$, \FBwd \;
}
\end{small}
\end{algorithm2e}
\begin{algorithm2e}[tbh]
\caption{\revise{Circulant-Sparse Attention {\small $\mathcal{O}(QW^2M)$}}}
\label{alg:circatt}
\DontPrintSemicolon
\begin{small}
\SetKwFunction{FFwd}{CircAtt}
\SetKwProg{Pn}{function}{:}{\KwRet}
\SetKwFunction{FBwd}{bwd}
\Pn{\FFwd{$\ADJMAT \in \BCCB_W^{Q\times Q}$, $\x \in \C^{QM}$}}{
    $\y = \cwise{\ADJMAT}\x$ \;

    \BlankLine

    \Pn{\FBwd{$\dd\y \in \C^{QM}$}}{
        $\dd{\ADJMAT} =
        \begin{cases}
            \dd\y[j]^T\x[i], &
            \norm{\vec{i} - \vec{j}}_\infty \leq \tfrac{1}{2}\WINSZ \\
            0 , & \text{otherwise}
        \end{cases} {\textstyle \forall ~ (i,j) \in [1,Q]^2}$\; 
        $\dd{\x} = \cwise{\ADJMAT}^T \dd\y$\;
        \KwRet $\dd{\ADJMAT}$, $\dd{\x}$\;
    }
    \KwRet $\y$, \FBwd \;
}
\end{small}
\end{algorithm2e}
\subsection{Circulant-Sparse Attention} \label{sec:circatt}
\revise{GroupCDL approaches the nonlocal self-similarity prior from the perspective of
group-sparsity of convolutional sparse-codes. This perspective dictates the
formation and use of similarity matrices which encode a group-membership
relationship between all pixels of an input image.
Ideally, latent pixels from anywhere in the image, regardless of proximity,
may belong a group if they represent similar image regions. 
However, for computational and memory reasons, it is infeasible to form such dense
matrices even for reasonably small image sizes\footnote{
    Memory and computation scale with side-length to the fourth power. A dense
    Float32 adjacency matrix for a square images of side lengths 256 and 512
    would require 17 and 275 GB of memory, respectively
}. 
It is thus appropriate to spatially limit our computations to more reasonably sized
windows. We seek to limit similarity computations for the $i$th pixel
to pixels $j$ within a window of side-length $W$, i.e. for all $j$ such that $\norm{\vec{i}
- \vec{j}}_\infty \leq \frac{1}{2}\WINSZ$ where $\vec{\cdot}$ notation denotes the conversion
from linear to Cartesian indexing (see Table \ref{tab:notation}). The result of which is a sparse 
similarity matrix.}

\revise{Previous nonlocal DNN architectures have managed computational complexity in a
different way: by independently processing overlapping image patches for the
entire network \cite{lecouat2020nonlocal, liu2018non}, or latent patches in
each layer \cite{Liang_2021_ICCV}.}
After this processing, image/latent patches 
may be combined to re-form the image/latent via averaging patches overlapping regions. 
\revise{We illustrate this ``Patch-based Dense Attention" (PbDA) in Figure \ref{fig:HT:dense}.}
This patching operation follows arbitrary borders and necessarily
neglects dependencies between pixels of neighboring patches, despite their
proximity.

In GroupCDL we propose a return to the original goal of a
spatially-limited nonlocal attention. This goal is achieved by three
observations. First, spatially-limited attention defines a Block-Circulant with
Circulant Blocks (BCCB) sparsity pattern for the adjacency matrix (when
circular boundary conditions are imposed)\footnote{This is the sparsity pattern
of a 2D single-channel convolution matrix}, as seen illustrated in Figure
\ref{fig:HT:circ}. As each row of the BCCB-sparse matrix (with windowsize
$W\times W$) contains $W^2$ elements, the non-zero elements are easily arranged
in a matrix $\tilde{\bS} \in \R^{N \times W^2}$. This sparse matrix may be
populated with similarity computations in parallel, and a row-normalization may
be performed on $\tilde{\bS}$ using existing fast softmax algorithms.

Second, the now formed BCCB-sparse adjacency matrix may be applied to a vectorized image 
using existing fast sparse-dense matrix multiplication algorithms.

Third, the back-propagation rules for BCCB-sparse similarity computations and
BCCB-sparse attention make use of and maintain the original BCCB sparsity
pattern. Memory explosion from considering a $N\times N$ adjacency matrix is
not possible, as the sparsity pattern remains fixed. This allows for training
with circulant-sparse attention, enabling effective tackling of image
restoration with global artifacts/degradation operators, as explored in Section
\ref{sec:exp:csmri}.

The proposed GroupCDL makes use of a circulant-sparse distance similarity and
circulant-sparse attention, detailed in Algorithms \ref{alg:circdist},
\ref{alg:circatt}. In each of these algorithms, a backward-pass function is
returned which produces the input gradients (ex. $\dd{\bk}, \dd{\bq},
\dd{\ADJMAT}, \dd{\x}$) given output gradients (ex. $\dd\bS, \dd\y$). \revise{The time
complexity of each algorithm is given next to their respective names. Note that
the backward passes and forward passes of each algorithm share the same
complexity. For latent images with $Q$-pixels and $M$-channels, and a
windowsize of $W\times W$, this complexity is $\mathcal{O}(QW^2M)$. This is in
stark contrast to naive dense attention, which scales quadratically with pixels
via $\mathcal{O}(Q^2M)$. Further comparison between CircAtt and dense attention
is given in Section \ref{sec:slidingwindow}.}

Derivations of these back-propagation rules are given in the Supplementary Material. 
Our Circulant-Sparse Attention is implemented in Julia's
\href{https://github.com/JuliaGPU/CUDA.jl}{CUDA.jl} \cite{besard2018juliagpu} and is publicly
available\footnote{\href{https://github.com/nikopj/CirculantAttention.jl}{https://github.com/nikopj/CirculantAttention.jl}}.

\subsection{Circulant-Sparse vs. Patch-based Dense Attention} \label{sec:slidingwindow}
\revise{In this Section, we discuss the benefit of CircAtt
vs. PbDA. We consider these methods orthogonal to
the function forms of attention (see Section \ref{sec:dpa}) as they by and
large concern the formation of their respective adjacency matrices. In
principle, one may use PbDA with GroupCDL instead of CircAtt.}

\revise{The proposed CircAtt, employed by GroupCDL, is favorable to PbDA, employed by
GroupSC \cite{lecouat2020nonlocal} and other black-box DNNs \cite{liu2018non, Liang_2021_ICCV, Valsesia2020}, because it naturally encourages agreement on
overlapping regions and centers the attention windows on each pixel. As shown
in Figure \ref{fig:HT:dense}, PbDA incurs computational overhead by processing
overlapping pixels multiple times.} Let $N = N_1 \times N_2$ be the number of
pixels in an image processed by window-size $\WINSZ \times \WINSZ$ (for both
PbDA and CircAtt), and denote window-stride $\WINSTRIDE \times \WINSTRIDE$. 
We express the burden factor as a ratio of the number of pixels processed by a
single PbDA-enabled nonlocal layer over a CircAtt-enabled nonlocal layer,
\begin{equation} \label{eq:burden}
    \frac{N_1/\WINSTRIDE \times N_2/\WINSTRIDE \times \WINSZ^2}{N_1 \times N_2} = \frac{\WINSZ^2}{\WINSTRIDE^2}.
\end{equation}
Common nonlocal window sizes, $45 \times 45$,  and window-strides, $7
\times 7$, such as used by NLRN \cite{liu2018non}, make this burden factor 41 times the
computational complexity of an equivalent CircAtt implementation. GroupCDL's use of strided
convolution may add an additional $s_c^2 \times$  computational benefit compared
to common NLSS implementations by computing similarities over a reduced spatial
dimension $Q = N/s_c^2$. 

Note that shifted-window (Swin) attention (employed by SwinIR
\cite{Liang_2021_ICCV}) side-steps this burden factor by patching without
overlap ($\WINSTRIDE = \WINSZ$) twice (successively) with shifted patch
boundaries. Swin attention may also enjoy enhanced processing speeds due to its
use of dense-array arithmetic vs. CircAtt's use of sparse-array arithmetic.
Yet, Swin attention does not address the modeling deficiencies inherent with
patch-based dense attention. This is demonstrated in experimentally by
GroupCDL's on-par performance with SwinIR at a fraction of its learned
parameter count, in Section
\ref{sec:exp:single}.

\revise{We further explore the relation between computation time and denoising
performance of PbDA vs CircAtt in the Supplementary Material.}

\subsection{Group-Thresholding vs. Black-box Attention} \label{sec:dpa}
Nonlocal self-similarity is used across domains in DNNs, from transformer
architectures \cite{dao2022flashattention} to nonlocal image restoration networks
\cite{liu2018non, Liang_2021_ICCV}. The underlying formula behind these
methods is most commonly dot-product attention (DPA), easily expressed in terms of the reshaped latent 
$\bZ \in \R^{N\times M_{\mathrm{in}}}$ as,
\begin{equation} \label{eq:dotprod_atten_2}
    \begin{gathered} 
        \bK = \bZ^{(k)}\bW_k^T, \quad \bQ = \bZ^{(k)}\bW_q^T, \quad \bV = \bZ^{(k)}\bW_v^T  \\
        \ADJMAT = \rowsoftmax(\bK\bQ^T) \\
        \bZ^{(k+1)} = \ADJMAT \bV
    \end{gathered}
\end{equation}
where $\bW_q, \bW_k, \in \R^{M_{h} \times M_{\mathrm{in}}}$,
$\bW_v \in \R^{M_{\mathrm{out}} \times M_{\mathrm{in}}}$, and $\bZ^{(k+1)} \in \R^{N \times M_{\mathrm{out}}}$. 
\revise{Here we refer to DPA as the functional form of black-box attention, orthogonal
to the (implicit) formation of the entire-image's adjacency matrix which is by
and large dictated by the use of PbDA for
black-box networks (see Section \ref{sec:slidingwindow}). This comes from the
fact that \eqref{eq:dotprod_atten_2} computes a dense $\ADJMAT$, which is
infeasible for use on an entire image. In principle, one may use DPA with
CircAtt instead of with PbDA.}

Both DPA and the proposed $\GT$ \eqref{eq:GTlearned} make use of a normalized adjacency matrix
($\ADJMAT$), computed in an asymmetric feature domain\footnote{($\bW_k \neq \bW_q$, $\bW_\theta \neq \bW_\phi$)}. Both use this adjacency
to weight the current spatial features, identically over channels. However, in
DPA, the weighting directly results in the layer's output (via matrix
multiplication), whereas in $\GT$ this weighting informs a spatially-varying
image-adaptive soft-thresholding. 

Our proposed $\GT$'s decoupling of adjacency matrix application and layer output dimension is key in allowing
learned group-thresholding to be computationally efficient, as the adjacency
matrix-vector multiplication can be performed over transformed features with a reduced dimension. 
In contrast, DPA operating with reduced channel dimensions ($M_{\mathrm{out}}
<< M_{\mathrm{in}}$) would harm the capacity of the network's latent
representation. 

\section{Experimental Results} \label{sec:results}
\subsection{Natural Image Denoising Experimental Setup}
\textbf{Architecture}: GroupCDL and CDLNet are trained with noise-adaptive
thresholds ($\btau^{(k)} = \btau^{(k)}_0 + \hat{\sigma}\btau^{(k)}_1$) unless
specified using the -B suffix, indicating the models are noise-blind
($\btau^{(k)} = \btau^{(k)}_0$). The hyperparameters for these architectures
are given in Table \ref{tab:arch}, unless otherwise specified. 

\begin{table}[tb]
\caption{Architectures of the GroupCDL models, CDLNet models, and variants presented in the experimental section. 
Conv-stride $s_c=2$ for all models.
}
\centering
\resizebox{\linewidth}{!}{%
\begin{tabular}{cccccccc} \hline
    Name & Task & $p$ & $K$ & $M$ & $M_h$ & $W$ & $\Delta K$ \\ \hline
     CDLNet & Denoise       & 7 & 30 & 169 & -   & - & - \\ 
     GroupCDL & Denoise, CS-MRI & 7 & 30 & 169 & 64  & 35 & 5 \\ 
     Big-GroupCDL & Denoise & 9 & 40 & 448 & 128 & 45 & 10 \\ 
     \hline
\end{tabular}
}
\label{tab:arch}
\end{table}

\textbf{Dataset and Training}:
Let $f_\Theta$ denote the GroupCDL DNN as a function of
parameters $\Theta$. Let $\D = \{(\y, \sigma, \x)\}$ denote a dataset of noisy and
ground-truth natural image pairs, with noise-level $\sigma$. Grayscale image denoising GroupCDL models were trained on
the BSD432 \cite{bsd} dataset. Grayscale Big-GroupCDL models were additionally trained on the Waterloo Exploration dataset and DIV2K dataset. All models were trained with a supervised mean squared
error (MSE) loss,
\begin{equation} \label{eqn:mse}
\underset{
\substack{
\bW_\theta, ~ \bW_\phi,~\bW_\alpha,\\
\bW_\beta \geq 0,  ~ \gamma \in [0, 1], \\
\bm{D} \in \mathcal{C}, ~ \{ \brho^{(k)}, \, \btau^{(k)} \geq 0 \}_{k=0}^{K-1}, \\
\{ \bm{A}^{(k)} \in \mathcal{C}, ~ \bm{B}^{(k)} \in \mathcal{C} \}_{k=0}^{K-1}
}
}{\mathrm{minimize}} \quad \sum_{\{\y, \sigma, \x\} \in \mathcal{D}} 
\norm{f_{\Theta}(\y, \sigma) - \x}_2^2,
\end{equation}
where $\mathcal{C} = \{ \bD \, : \, \norm{\bD_{:j}}_2^2 \leq 1 ~
\forall \, j \}$. We use the Adam
optimizer with default parameters \cite{adam}, and project the network parameters
onto their constraint sets after each gradient step.
The dataset is generated online from clean images via random crops, rotations, flips, and
AWGN of noise-level $\sigma$ sampled uniformly within $\sigmatrain$ for each mini-batch element. 
A mini-batch size of 12 was used. Initial learning-rates of $5\times 10^{-4}$ and $3\times 10^{-4}$ were used for 
GroupCDL and Big-GroupCDL, respectively, with cosine-annealing \cite{loshchilov2016sgdr} to a final value of $2\times 10^{-6}$. 
Each network was trained for $\approx 600k$ gradient steps.
Additional training hyperparamter details (such as back-tracking) follow the
CDLNet setup \cite{janjusevicCDLNet2022}. 

Test and validation performance is evaluated on several datasets. The dataset name, along with (arithmetic) average 
dimensions, are provided to better understand reported inference timings: Set12 (362 $\times$ 362), CBSD68 \cite{bsd} (481 $\times$ 321), Urban100 \cite{Urban100} (1030 $\times$ 751), and 
NikoSet10\footnote{see \href{https://github.com/nikopj/GroupCDL}{https://github.com/nikopj/GroupCDL}.} (1038 $\times$ 779).

\textbf{Training Initialization}:
GroupCDL models are initialized as ISTA\footnote{$(\bA^{(k)} = \bB^{(k)} = \bD$ $\forall$ $k$)} with $\btau_0 = 10^{-3}, ~ \btau_1 = 0$,
and a base dictionary $\bD$ that has been spectrally normalized. Details are
given in \cite{janjusevicCDLNet2022}. Pixel-wise transforms $\bW_{\{\theta, \phi, \alpha, \beta\}}$
are initialized with the same weights drawn from a standard uniform distribution and spectrally-normalized.
We initialize parameter $\gamma = 0.8$.

\textbf{Hardware}:
All models were trained on a single core Intel Xeon CPU with 2.90 GHz clock and
a single NVIDIA A100 GPU. GroupCDL training takes approximately 48 hours. 
For code-base compatibility reasons, inference timings for methods
(GroupCDL, GroupSC, NLRN) in Tables \ref{tab:graysingle} were determined by
running models an NVIDIA Quadro RTX-8000, whereas for other listed methods
(Big-GroupCDL, SwinIR, Restormer, DCDicl) we report NVIDIA A100 inference time.
\revise{Inference times of competing methods were determined by running the publically
available code from the respective authors.}

\subsection{Single Noise-Level Performance} \label{sec:exp:single}
In this section, we demonstrate competitive
denoising performance of the proposed GroupCDL. All models are trained on a
single noise-level and tested at the same noise-level ($\sigmatrain =
\sigmatest$). We compare GroupCDL to its fully convolutional counterpart
(CDLNet), 
patch-processing dictionary learning based nonlocal DNN (GroupSC) \cite{lecouat2020nonlocal},
and
\soa black-box DNNs \cite{Valsesia2020,liu2018non,Liang_2021_ICCV,Zamir2021Restormer,Zheng_2021_CVPR}.

\begin{table*}[ht]
\centering
\caption{
Grayscale denoising performance (PSNR (dB)/ $100\times$SSIM) and GPU inference
runtimes. All learned methods are trained for individual noise-levels ($\sigma
= \sigmatrain = \sigmatest$). Learned parameter counts are displayed below the
method names. \revise{Methods labeled with $^\star$ are nonlocal. Methods labeled with
$^\diamond$ are nonlocal and use PbDA. Column $\Delta$ represents the performance difference between the proposed Big-GroupCDL and SwinIR \cite{Liang_2021_ICCV}.}
}
\resizebox{\linewidth}{!}{%
\begin{tabular}{|cc|HHcccc|HHccccHc|} \hline
    Dataset & $\sigma$ & 
    \makecell{BM3D \\ - \cite{bm3d}} & 
    \makecell{DnCNN \\ 556k \cite{DnCNN}} &
    \makecell{CDLNet \\ 507k \cite{janjusevicCDLNet2022}} & 
    \makecell{GroupSC$^\diamond$ \\ 68k \cite{lecouat2020nonlocal}} & 
    \makecell{NLRN$^\diamond$ \\ 340k \cite{liu2018non}} & 
    \makecell{GroupCDL$^\star$ \\ 550k} &
    \makecell{GCDN$^\diamond$ \\ 6M \cite{Valsesia2020}} & 
    \makecell{MWCNN \\ 14M \cite{Liu2018}}  & 
    \makecell{SwinIR$^\star$ \\ 11M \cite{Liang_2021_ICCV}} & 
    \makecell{Restormer$^\star$ \\ 25M \cite{Zamir2021Restormer}} & 
    \makecell{DCDicL \\ 33M \cite{Zheng_2021_CVPR}} & 
    \makecell{Big-GroupCDL$^\star$ \\ 3M} &
    $\Delta$ & 
    $\Delta$   
    \\\hline
\multirow{3}{*}{Set12} 
& 15       & 32.37/89.52 & 32.86/90.31 & 32.87/90.43 & 32.85/90.63 & {\bf 33.16}/\underline{90.70} & \underline{33.10}/{\bf 90.81} & 33.14/90.72 & 33.15/90.88 & \underline{33.36}/91.11 & {\bf 33.42}/{\bf 91.28} & 33.34/\underline{91.15} & 33.24/91.02 & 0.10/0.13 & -0.12/-0.09 \\
& 25       & 29.97/85.04 & 30.44/86.22 & \underline{30.52}/86.55   & 30.44/86.42                   & {\bf 30.80}/\underline{86.89} & {\bf 30.80}/{\bf 87.12} & 30.78/86.87 & 30.79/87.11 & 31.00/87.38 & {\bf 31.08}/{\bf 87.64} & \underline{31.03}/\underline{87.48} & 30.91/87.35 & 0.14/0.13 & -0.09/-0.03 \\
& 50       & 26.72/76.76 & 27.18/78.29 & 27.42/79.41 & 27.14/77.97 & \underline{27.64}/\underline{79.80} & {\bf 27.73}/{\bf 80.52} & 27.60/79.57 & 27.74/80.56 & 27.87/80.84 & \underline{27.95}/\underline{81.18} & {\bf 28.00}/{\bf 81.22} & 27.85/80.82 & 0.10/0.36 & -0.02/-0.02 \\
time (s) & & 0.010       & 0.119       & 0.019       & 22.07       & 25.62       & 0.68            & 405         &             & 1.88        & 0.13        & 0.42        & 1.60    &  &  -0.18 \\\hline
                                                                                                                                                                                    
\multirow{3}{*}{\makecell{BDS68 \\ \cite{bsd}}}                                                
 & 15       & 31.07/87.17 & 31.73/89.07 & 31.74/89.18 & 31.70/{\bf 89.63} & {\bf 31.88}/89.32 & \underline{31.86}/\underline{89.49} & 31.70/\underline{89.63} & 31.86/89.47 & {\bf 31.96}/\underline{89.60} & {\bf 31.96}/{\bf 89.65} & \underline{31.95}/89.57 & 31.90/89.57 & 0.05/0.03 & -0.06/-0.03 \\
& 25       & 28.57/80.13 & 29.23/82.78 & 29.26/83.06 & 29.20/\underline{83.36} & \underline{29.41}/83.31 & {\bf 29.42}/{\bf 83.60} & 29.35/83.32 & 29.41/83.60 & \underline{29.50}/83.77 & {\bf 29.52}/{\bf 83.90} & {\bf 29.52}/\underline{83.79} & 29.46/83.76 & 0.04/0.03 & -0.04/-0.01 \\
& 50       & 25.62/68.64 & 26.23/71.89 & 26.35/72.69 & 26.17/71.83 & \underline{26.47}/\underline{72.98} & {\bf 26.51}/{\bf 73.63} & 26.38/73.89 & 26.53/73.66 & 26.57/73.79 & \underline{26.60}/{\bf 73.99} & {\bf 26.63}/\underline{73.95} & 26.57/73.83 & 0.03/0.12 & +0.0/+0.04 \\
time (s) & & 0.011       & 0.039       & 0.022       & 23.63       & 26.66       & 0.65        & 540         &             & 1.43        & 0.14        & 0.38        & 1.67   &   &  +0.24 \\\hline
                                                                                                                                                                                    
\multirow{3}{*}{\makecell{Urban100 \\ \cite{Urban100}}}                                         
& 15       & 32.35/92.20 & 32.68/92.55 & 32.59/92.85 & 32.72/93.08 & {\bf 33.42}/\underline{93.48} & \underline{33.24}/{\bf 93.58} & 33.47/93.58 & 33.17/93.57 & \underline{33.75}/\underline{93.98} & {\bf 33.79}/{\bf 94.09} & 33.59/93.88 & 33.56/93.89 & 0.19/0.09 & -0.19/-0.09 \\
& 25       & 29.70/87.77 & 29.92/87.97 & 30.03/89.00 & 30.05/89.12 & {\bf 30.88}/\underline{90.03} & \underline{30.39}/{\bf 90.18} & 30.95/90.20 & 30.66/90.26 & \underline{31.30}/90.94 & {\bf 31.46}/{\bf 91.29} & \underline{31.30}/\underline{91.08} & 31.20/90.93 & 0.10/0.15 & -0.10/-0.01 \\
& 50       & 25.95/77.91 & 26.28/78.74 & 26.66/81.11 & 26.43/80.02 & {\bf 27.40}/\underline{82.44} & \underline{27.29}/{\bf 83.64} & 27.41/81.60 & 27.42/83.71 & 28.05/84.84 & {\bf 28.29}/{\bf 85.60} & \underline{28.24}/\underline{85.49} & 28.03/84.83 & 0.21/0.67 & -0.02/-0.01 \\
time (s) & & 0.030       & 0.096       & 0.090       &  93.33      & 135.8       & 3.56        & 1580        &             & 4.98        & 0.60        & 1.75        & 8.32   &  &  +3.34 \\\hline
\end{tabular}
}
\label{tab:graysingle}
\end{table*}
Table \ref{tab:graysingle} shows the grayscale
denoising performance and inference speed of the aforementioned models across
several common datasets. We
include a learned parameter count as a crude measure of expressivity of the model.
The group-sparsity prior of GroupCDL significantly increases
denoising performance compared to the unstructured sparsity prior of CDLNet.
We observe that GroupCDL has denoising performance superior to other dictionary
learning based networks using group sparsity prior (GroupSC) and competitive performance with 
the black-box nonlocal method NLRN. GroupCDL also significantly outperforms these methods
in inference time due to the computational efficiency of CircAtt over PbDA.
Big-GroupCDL is shown to be competitive with \soa black-box networks SwinIR, Restormer, and DCDicl, with 
on par inference time and only a fraction of the learned parameters.

\begin{figure*}[thb]
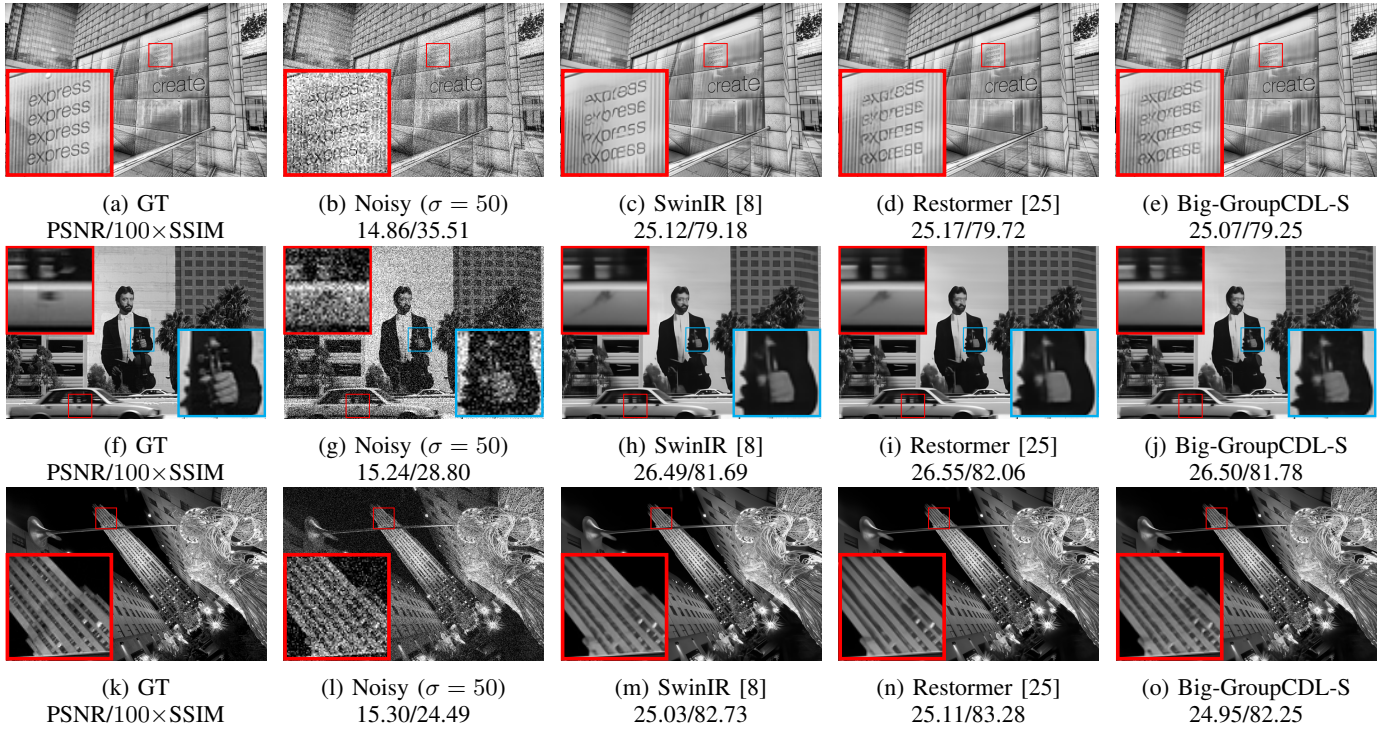

    \centering
    \captionsetup[subfigure]{justification=centering}
    \input{figs/compare_BigGrayS_Express.tex}
    \\
    \input{figs/compare_BigGrayS_bsd.tex}
    \\
    \input{figs/compare_BigGrayS_urban.tex}
    \caption{
    Visual comparison of large paramter count deep denoisers. 
    }
    \label{fig:biggray}
\end{figure*}

Figure \ref{fig:biggray} highlights the qualitative differences between
Big-GroupCDL and \soa black-box networks (SwinIR \cite{Liang_2021_ICCV}, Restormer \cite{Zamir2021Restormer}). We observe that
Big-GroupCDL is able to produce denoised images on par with SwinIR and
Restormer (Fig. \ref{fig:biggray} first row red box), and may even retain a greater amount of detail information where
SwinIR and Restormer appear to over-smooth detailed regions (Fig. \ref{fig:biggray} second row red box, third row red box) or hallucinate image structures (Fig. \ref{fig:biggray} second row blue box). 

\subsection{Noise-Level Generalization}
\begin{figure}[tbh]
    \centering
    \begin{subfigure}{0.53\columnwidth}
        \includegraphics[width=\columnwidth]{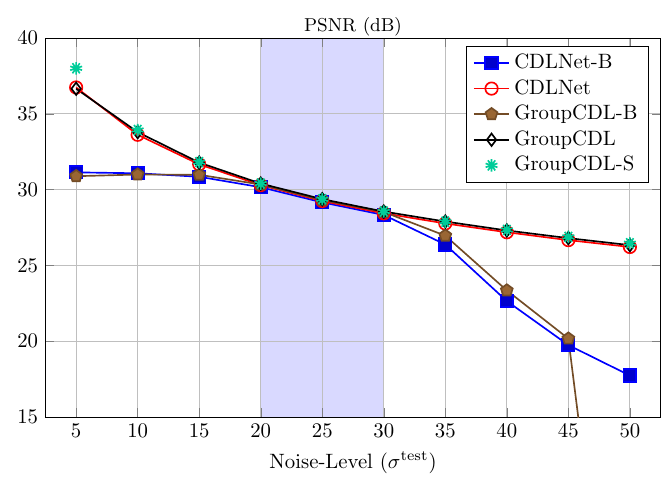}
        \vspace{-2.3em}
    \end{subfigure}%
    \hfill%
    \begin{subfigure}{0.46\columnwidth}
        \includegraphics[width=\columnwidth]{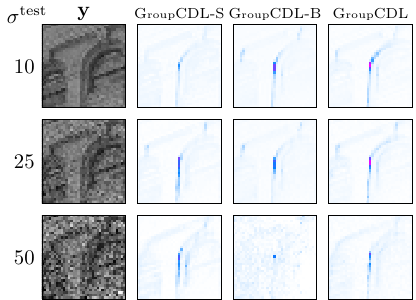}
    \end{subfigure}
    \caption{
        Left: Noise-level generalization of different grayscale denoising
        networks tested on BSD68 \cite{bsd}. GroupCDL-S is trained at
        $\sigmatest$ for each point on the graph. All other networks are
        trained on $\sigmatrain =[20, 30]$.
        Right:
        Visualization of normalized adjacency $\ADJMAT^{(K)}_{i:}$ for $i$th
        pixel of input image $(\y)$ at noise-levels $\sigma=10,25,50$. 
        Catastrophic failure of GroupCDL-B model is
        observed for $\sigma > \sigmatrain$. 
    }
    \label{fig:adj_gen}
\end{figure}%
Figure
\ref{fig:adj_gen} shows that the proposed novel group-thresholding
scheme \eqref{eq:GTlearned} is able to obtain near-perfect noise-level
generalization (w.r.t GroupCDL-S performance). 
This serves as empirical evidence for
the interpretation of the unrolled network as performing some
approximate/accelerated group-sparse BPDN, as the noise-adaptive thresholds
($\btau = \btau_0 + \hat{\sigma}\btau_1$; $\hat{\sigma}$ estimated following \cite{janjusevicCDLNet2022}) appear to correspond very well to
their classical counter-parts from which they are derived. 

Figure \ref{fig:adj_gen} also shows a single input nonlocal
window $\y$ across noise-levels $\sigma$ and the computed adjacency values of
the three types of GroupCDL models (GroupCDL-S, GroupCDL-B, GroupCDL). 
The adjacency visualizations of
GroupCDL-B show a catastrophic failure in the
similarity computations of the network above $\sigmatrain$, as no structure is
found. Similar patterns are seen in the
GroupCDL models (with noise-adaptive thresholds) compared to those of 
GroupCDL-S.

\subsection{Ablation of Learned Group-Thresholding} \label{sec:exp:ablation}
\revise{In this section we examine the denoising and inference time performance of the
GroupCDL model under different hyperparameters associated with the proposed
learned group-thresholding operation \eqref{eq:GTlearned}. }
\begin{table}[ht]
    \centering
    \caption{\revise{Effect of learned transforms in Learned Group-Thresholding. Grayscale
    denoising performance averaged over the NikoSet10 dataset
    ($\sigmatrain=\sigmatest=25$). GroupCDL-S is used for all tests with
    $M=169$.}}
    \begin{tabular}{cc|cc}
        \hline
        learned transforms & $M_h$ & PSNR/$100\times$SSIM & time (s) \\
        \hline
        none & n/a & 30.19/86.79 & 8.64 \\
        $\bW_{\{\theta, \phi\}}$ & 64 & 30.21/86.86 & 8.13 \\
        $\bW_{\{\theta, \phi, \alpha, \beta\}}$ & 169 & 30.21/86.89 & 8.89 \\
        $\bW_{\{\theta, \phi, \alpha, \beta\}}$ & 64 & 30.21/86.87 & 3.47 \\
        $\bW_{\{\theta, \phi, \alpha, \beta\}}$ & 32 & 30.20/86.84 & 2.00 \\
        \hline
    \end{tabular}
    \label{tab:ablation:nlss}
\end{table}
\revise{Table \ref{tab:ablation:nlss} shows the effect of employing learned pixel-wise
transforms in the similarity ($\bW_{\theta,\phi}$) and attention
($\bW_{\alpha,\beta}$) computations of the proposed learned group-thresholding
\eqref{eq:GTlearned}. The table also shows the effect of employing channel
reduction in these transforms ($M_h << M = 169$). }
\revise{We observe that pixel-wise transforms only marginally increase denoising
performance, and are not integral to GroupCDL's performance. However, setting
$M_h << M$ greatly reduces inference time,  with the reduction roughly equal to
the channel reduction ratio $M / M_h = 169/64 \approx 8.89/3.47$, as suggested
by the time-complexities listed in Algorithms \ref{alg:circdist} and
\ref{alg:circatt}.} This demonstrates one advantage of learned
group-thresholding over black-box dot-product attention: the dimension for
similarity calculation and attention operation ($M_h$) is decoupled from the
layer's output channel dimension and may be tuned to achieve a better trade-off
between speed and performance. 

\newcommand{\eightx}{8$\times~$}
\newcommand{\fourx}{4$\times~$}
\subsection{Application to Compressed-Sensing MRI} \label{sec:exp:csmri}
\begin{table*}[tb]
\centering
\caption{CS-MRI reconstruction performance (PSNR (dB)/$100\times$SSIM. All learned methods 
    are trained on the MoDL Brain dataset \cite{Aggarwarl2019_TMI}. Learned
    parameter counts are displayed below the method names. $^\dagger$Numbers reported in \cite{Liu2023_TCI}.
    \revise{Methods labeled with $^\star$ are nonlocal. 
    Note that the zero-filled reconstruction metrics closely match those reported in \cite{Liu2023_TCI}.}
    }
    \label{tab:csmri}
\resizebox{\linewidth}{!}{%
\begin{tabular}{ccccccccc} \hline
    \multirow{2}{*}{Accel.} & Zero-Filled & MoDL$^\dagger$ \cite{Aggarwarl2019_TMI} & ISTA-Net$+$$^\dagger$ \cite{Zhang2018_CVPR} & VS-Net$^\dagger$ \cite{Duan2019_MICCAI} & E2EVarNet \cite{sriram2020end} & HFMRI$^\dagger$$^\star$ \cite{Liu2023_TCI} & CDLNet & GroupCDL$^\star$  \\
    ~ & - & 5.6M & 368k & 1.0M & 30M & 1.1M & 1.0M & 1.1M \\\hline
    4x & 25.60/80.61 & 28.89/80.1 & 32.26/91.6 & 32.18/91.4 & 31.2/91.4 & 33.20/93.8 & \underline{36.3}/\underline{96.2} & {\bf 36.6}/{\bf 96.4} \\
    8x & 23.24/72.91 & 25.28/71.1 & 27.48/83.1 & 27.56/83.3 & 25.6/81.0 & 27.65/85.3 & \underline{29.3}/\underline{88.1} & {\bf 30.5}/{\bf 89.9}  \\
    \hline 
\end{tabular}}
\end{table*}
\begin{figure*}[thb]
    \centering
    \captionsetup{justification=centering,singlelinecheck=false}
    \input{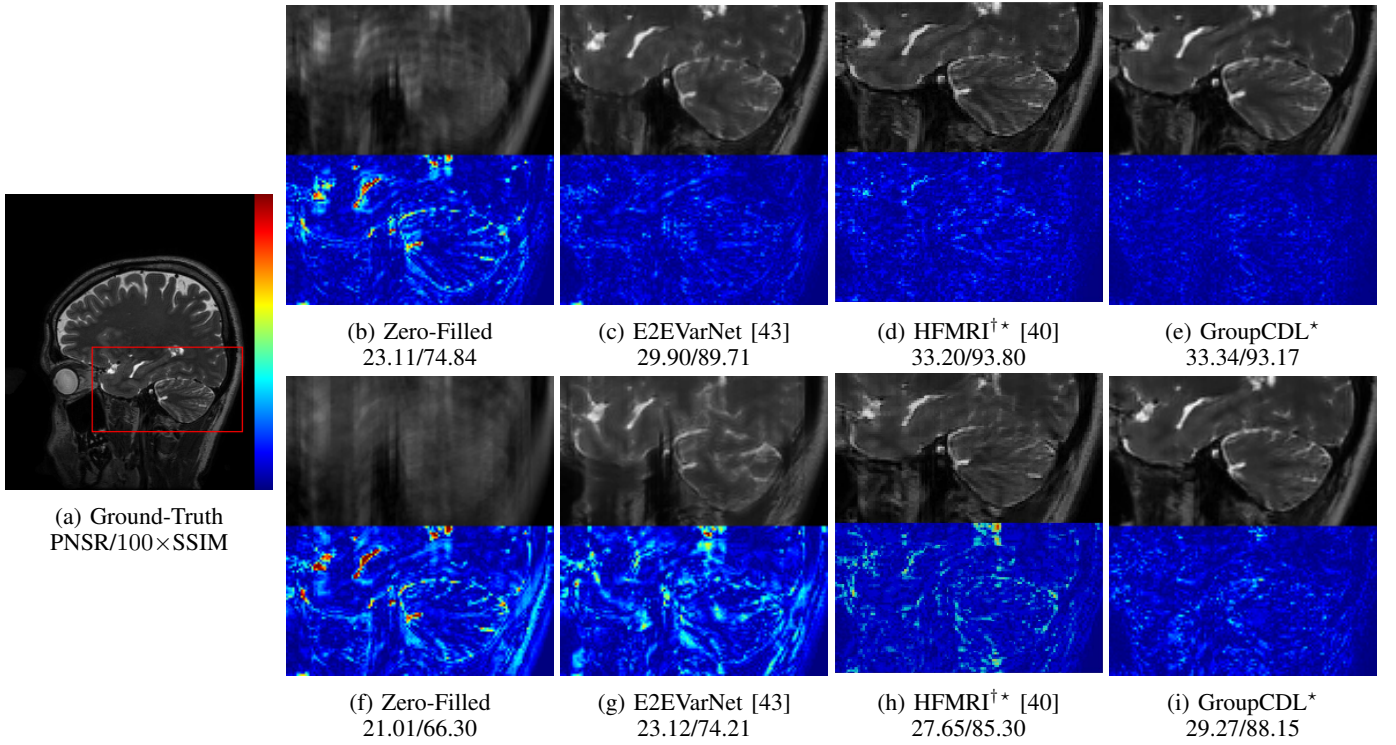}
    \captionsetup{justification=justified,singlelinecheck=false}
    \caption{
        Visual comparison of multi-coil CS-MRI reconstruction models on a selected crop of slice
        111 from the MODL Brain test set \cite{Aggarwarl2019_TMI}. 
        Magnitude error maps shown below each crop.
        Top-row: 4x acceleration. Bottom-row: 8x acceleration. $^\dagger$figure and
        numbers taken from \cite{Liu2023_TCI}. Methods labeled with $^\star$ are nonlocal. 
    }
    \label{fig:modl_zero}
\end{figure*}
\begin{figure}[thb]
    \centering
    \captionsetup{justification=centering,singlelinecheck=false}
    \begin{subfigure}{0.33\columnwidth}
        \centering
        \includegraphics[width=\linewidth]{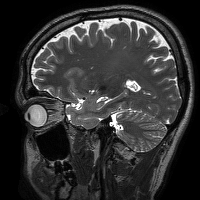}
        \caption{GT \\ PNSR/$100\times$SSIM}
    \end{subfigure}%
    \hfill%
    \begin{subfigure}{0.65\columnwidth}
        \centering
        \includegraphics[width=\linewidth]{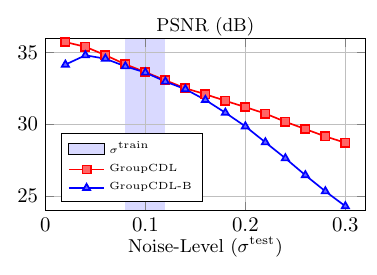}
        \vspace{-2em}
        \caption{}
        \label{fig:modl_gen_plot}
    \end{subfigure}%
    \\
    \begin{subfigure}{\linewidth}
        \input{figs/modl_gen/figure.tex}
    \end{subfigure}
    \captionsetup{justification=justified,singlelinecheck=false}
    \caption{
        Visual comparison of Joint-Denoising and CS-MRI reconstruction models
        with (GroupCDL) and without noise-adaptive thresholds (GroupCDL-B) on
        MODL Brain test set \cite{Aggarwarl2019_TMI}. Each model was trained for \fourx acceleration on a noise-range 
        $\sigmatrain \in [0.08, 0.12]$. 
        Top-row: $\sigmatest = 0.02$. 
        Bottom-row: $\sigmatest = 0.25$. 
    }
    \label{fig:modl_gen}
\end{figure}
In this section, we apply the GroupCDL architecture (detailed in Section
\ref{sec:csmri}) to supervised training for the CS-MRI problem, with and without noise
contamination. \\
\noindent \textbf{Experimental Setup}: We use the fully-sampled MoDL multi-coil brain dataset
\cite{Aggarwarl2019_TMI}, which was acquired using a 3D T2 CUBE sequence with
an isotropic 1 mm resolution, a 210 mm $\times$ 210 mm field-of-view, and
echo-time of 84 ms. The dataset contains 360 slices for training and 164 for
testing. Sensitivity maps are provided by the dataset. We follow the
experimental setup of HFMRI \cite{Liu2023_TCI} by using random Cartesian Fourier domain
subsampling (in a single readout direction) for \fourx and \eightx
acceleration with $8\%$ and $4\%$ of center lines preserved, respectively. \\
\textbf{Training}: \revise{We train CDLNet and GroupCDL models for CS-MRI 
using image crop-sizes of $200\times 200$ and otherwise identical model and
training hyperparameters as the grayscale denoising models (see Table
\ref{tab:arch}), though each filter is now complex-valued (nearly doubling the
learned parameter count).} Following \cite{Liu2023_TCI}, we replace the standard
MSE loss function with an $\ell_1 - \mathrm{ssim}$ loss function for fair
comparison. PSNR and SSIM computations use a peak-value of $1.0$, following
\cite{Liu2023_TCI}. \revise{We verify the correctness of our implementation by
comparing the metrics of our zero-filled reconstructions (Table
\ref{tab:csmri}) with those reported in HFMRI \cite{Liu2023_TCI}. We
additionally train an implementation of E2EVarNet \cite{sriram2020end} based
directly on their publically available code\footnote{see \href{https://github.com/facebookresearch/fastMRI}{https://github.com/facebookresearch/fastMRI}.}.} \\
\noindent \textbf{No Noise}: 
Table \ref{tab:csmri} shows the results of training the proposed GroupCDL
against \soa deep-learning CS-MRI reconstruction, without the presence of
additive noise. \revise{Both GroupCDL and CDLNet perform significantly better than \soa
methods, including E2EVarNet \cite{sriram2020end} which uses an order of magnitude more learned parameters\footnote{We observed E2EVarNet's validation PSNR drop after an initial rise, while its training loss continued to decrease, suggesting that E2EVarNet's high parameter-count caused it to overfit on the small training dataset.}.} We note that GroupCDL distinguishes itself from CDLNet at the higher acceleration
reconstruction. \revise{These observations are bolstered in
Figure \ref{fig:modl_zero} where the black-box DNNs HFMRI \cite{Liu2023_TCI}
and E2EVarNet \cite{sriram2020end} are shown to have significantly more structure present in their
reconstruction error maps compared to GroupCDL.} \\
\noindent \textbf{With Noise}:
Figure \ref{fig:modl_gen} shows the results of CS-MRI reconstruction training
in the presence of Fourier domain AWGN, across noise-levels. The proposed
CS-MRI GroupCDL model was trained for \fourx reconstruction on an 
intermediate noise-range ($\sigmatrain=[0.08,0.12]$)\footnote{with an
image-domain maximum magnitude of $1.6$}. 
Such a training may be encountered in 
practice where ground-truth images are unavailable and the limited available data is corrupted with varying noise-levels.
Figure \ref{fig:modl_gen_plot} shows that GroupCDL's noise-adaptive thresholds
enable effective inference both above and below the training noise-range.
Figure \ref{fig:modl_gen}'s visual examples show that GroupCDL-B is unable to
adequately recover features/textures below the training noise-range, and introduces severe
artifacts above it. In contrast, GroupCDL (with noise-adaptive thresholds) is
able to recover additional image details below the training range,
and generalize gracefully above it.

\section{Discussion and Conclusion}
In GroupCDL, we adapt the classical, patch-processing based, group-sparsity
prior to convolutional sparse coding (CSC) and apply it to the
direct-parametrization unrolling frame-work of CDLNet
\cite{janjusevicCDLNet2022}.
In doing so, we arrive at a sliding-window NLSS consistent with the CSC model
(CircAtt), which addresses the modeling pitfalls of patch-based dense
attention. We demonstrate competitive denoising performance to \soa black-box networks
with notable efficiency in the number of learned parameters.
We show that the CircAtt operation allows GroupCDL to be easily and effectively
extended to global degradation operator inverse-problems such as CS-MRI.
GroupCDL brings substantial gain over current \soa black-box nonlocal
networks for CS-MRI in the case of zero-noise reconstruction.

Additionally, GroupCDL inherits the noise-level generalization capabilities of
CDLNet \cite{janjusevicCDLNet2022}. This robustness to training-inference
mismatch may be of special interest to practical scenarios, where ground-truth
is unavailable and inference noise-levels may deviate from what is present in a
training dataset. We leverage GroupCDL's noise-adaptive thresholds
to recover image details especially well in such cases. To the best of our
knowledge, robustness to training-inference noise-level mismatch for CS-MRI has
not been considered in prior literature.

When scaled up, GroupCDL performs very competitively with \soa black-box DNNs,
even without incorporating existing \soa modeling choices such as
multi-resolution processing and transposed-attention. Future work may consider
additionally incorporating these features into the GroupCDL framework to enable
even better performance and inference speeds. We believe GroupCDL's
interpretability and robustness are well suited to tackle other large signal
reconstruction problems with nonlocal image-domain artifacts, especially in the
unsupervised learning regime.

\section*{Acknowledgments}
This work was supported in part by National Institute of Biomedical Imaging and Bioengineering (NIBIB) grants: R01EB030549, R01 EB031083 and R21 EB032917. 
The authors additionally thank NYU HPC and NYU Langone Health Ultra-Violet for their computing resources and
technical support. The authors are grateful to Che Maria Baez for her
linguistic revisions on a preliminary draft of this manuscript.

\ifCLASSOPTIONcaptionsoff
  \newpage
\fi
\bibliographystyle{IEEEtran}
\bibliography{IEEEabrv,references}

\end{document}